\documentclass{iopconfser}

\usepackage[numbers,sort&compress]{natbib}
\usepackage{tabularray}
\DefTblrTemplate{firsthead,middlehead,lasthead}{default}{
}
\DefTblrTemplate{firstfoot}{default}{
  \UseTblrTemplate{contfoot}{default}
  \UseTblrTemplate{caption}{default}
}
\DefTblrTemplate{middlefoot}{default}{
  \UseTblrTemplate{contfoot}{default}
  \UseTblrTemplate{capcont}{default}
}
\DefTblrTemplate{lastfoot}{default}{
  \UseTblrTemplate{note}{default}
  \UseTblrTemplate{remark}{default}
  \UseTblrTemplate{capcont}{default}
}
\NewTblrTheme{fancy}{
  \SetTblrStyle{caption-tag}{font=\bfseries\small}
}

\usepackage[nolist,nohyperlinks]{acronym}
\usepackage{graphicx}
\usepackage{caption}
\usepackage{subcaption}
\usepackage{float}
\usepackage[capitalise]{cleveref}
\crefname{table}{Tab.}{Tabs.}
\crefname{section}{Sec.}{Secs.}

\begin{document}

\title{Offset Finding of Beamline Parameters on the METRIXS Beamline at BESSY II Using Machine Learning}

\author{David Meier$^{1,3}$, Thomas Zeschke$^{1}$, Peter Feuer-Forson$^{1}$, Bernhard Sick$^{2,3}$, Jens Viefhaus$^{1}$ and Gregor Hartmann$^{1,3}$}

\affil{$^1$%
Helmholtz-Zentrum f\"ur Materialien und Energie GmbH, Albert-Einstein-Straße 15, 12489~Berlin, Germany
}
\vspace{10pt}
\affil{$^2$%
Intelligent Embedded Systems, University of Kassel, Wilhelmsh\"oher Allee 73, 34121~Kassel, Germany
}%
\vspace{10pt}
\affil{$^3$%
Artificial
Intelligence Methods for Experiment Design (AIM-ED), Joint Lab Helmholtz-Zentrum für Materialien und Energie, Berlin (HZB) and University of Kassel
}%

\email{david.meier@helmholtz-berlin.de}

\begin{abstract}
Beamline alignment is challenging as the beamline components must be set up ideally so that the rays follow the desired optical path. Automated methods using a digital twin allow for faster diagnostics and improved beam properties compared to manual tuning. We introduce an automated method of finding the offsets to improve this digital twin model. These offsets represent the unknown but constant differences between the beamline parameter positions as set up at the physical beamline and the corresponding parameter positions of its digital twin. Our method assumes the capability to execute precise relative movements with a known step size for these parameters, although the absolute position information is unknown. By combining the surrogate model with a global optimizer, we successfully determine offsets for 34 beamline parameters on a simulated METRIXS beamline at the BESSY II synchrotron radiation source in Berlin.
\end{abstract}

\section{Introduction}
\begin{acronym}[DIKW-hierarchy]\itemsep0pt
\acro{GPU}{Graphics Processing Unit}
\acro{BESSY II}{Berliner Elektronenspeicherring für Synchrotronstrahlung II}
\acro{GPU}{Graphics Processing Unit}
\acro{MSE}{Mean Squared Error}
\acro{RMSE}{Rooted Mean Squared Error}
\acro{MLP}{Multilayer Perceptron}
\acro{CPU}{Central Processing Unit}
\acro{METRIXS}{Momentum and Energy resolved Resonant Inelastic X-ray scattering at molecular Systems}
\end{acronym}
A synchrotron beamline consists of many components: mirrors, slits, gratings, crystals, and alignment units. Thus, many configurations must be tuned and optimized, which is time-consuming and requires expert knowledge. Because the beamline setup changes regularly, the beamline alignment must be carried out regularly up to daily. Thus, a one-click solution for automatically returning to the beamline’s optimal state within minutes instead of hours of manual tuning is desirable.

We need a \textit{digital twin}, a virtual representation of a real-world beamline to provide such an approach. For this digital twin, we use a raytracing software that takes beamline configurations as input to calculate footprints---point clouds of ray coordinates at the end of the beamline. The beam transmitted through the beamline consists of many rays with slightly different coordinates and trajectories. The raytracing software simulates the trajectories of these rays from the light source through all optical elements to the end of the beamline. The measurements carried out using a focus measurement chamber at the real-world beamline correspond to the calculated footprints of the digital twin. However, discrepancies between the real-world beamline and the digital twin arise due to manufacturing inaccuracies. Additionally, alignments of the beamline components are in millimeter to micrometer scale, while a typical beamline is at the scale of tens of meters, prohibiting precise adjustments of these components in the real world. We call these differences, which are fixed for all configurations, \textit{offsets}. Once the offsets are found, the digital twin---corrected by these offsets---can be used to optimize the beam properties to desired targets.
\begin{figure}[H]
 \begin{minipage}{\linewidth}
      \centering
      \begin{minipage}{0.49\linewidth}
          \begin{figure}[H]
              \includegraphics[width=\linewidth]{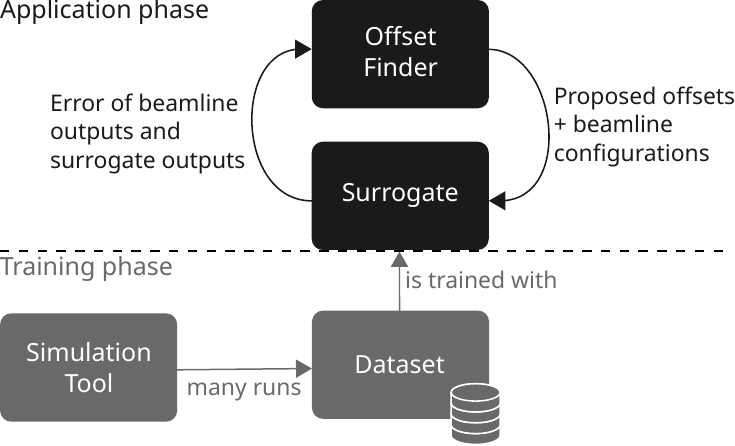}
              \caption{Architecture of our proposed approach.}
              \label{fig:graphical_abstract}
          \end{figure}
      \end{minipage}
      \hspace{0.001\linewidth}
      \begin{minipage}{0.465\linewidth}
          \begin{figure}[H]
              \includegraphics[width=\linewidth]{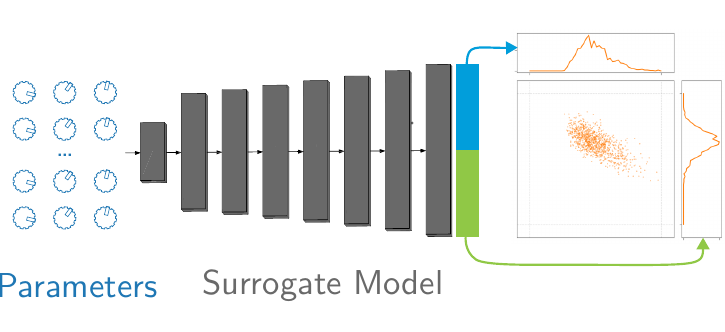}
              \caption{The beamline parameters are inputs to the neural network surrogate model, with histograms of the footprints as outputs.}
              \label{fig:params_to_hist}
          \end{figure}
      \end{minipage}
  \end{minipage}
\end{figure}
This study presents a method to find automatically the offsets for various beamline parameters. To gather sufficient information about the current parameter offsets, we manually search for multiple parameter configurations with many rays visible on the measurement chamber screen. This procedure is essential since only such configurations are expressive for offset finding. If few or no rays reach the end of the beamline, this footprint could result from many parameter configurations and thus provides insufficient information gain. An overview of the offset finding process is given in \cref{fig:graphical_abstract}. During the application phase, the offset finder suggests offsets, which are added to the previously determined beamline configurations with high ray counts. The configurations corrected by the proposed offsets must be simulated and compared with the footprints of the real-world beamline to evaluate the quality of the proposed offsets. Since the digital twin with 1-2 seconds evaluation time is too slow for this task, we replace it with a \textit{surrogate model}, a fast simulation approximation. This surrogate model is trained once during the training phase and can be repeatedly used for the offset-finding process in the application phase. For the surrogate model, we train a neural network with a dataset that contains many randomly selected combinations of the investigated beamline parameters and their results from the digital twin. To further accelerate the training and evaluation of the neural networks, we map the beamline configurations not to the footprints but their $x$ and $y$ histograms, which are binned projections along the $x$ and $y$ axes, as shown in \cref{fig:params_to_hist}. The offset finder uses the surrogate model to evaluate quickly the beamline configurations corrected by different proposed offsets, compares them to the outputs of the real-world beamline, and quantifies the errors. The offset finder then uses these errors to suggest new offsets to be evaluated. This process is repeated until the error between the beamline and surrogate model outputs becomes sufficiently small.

We demonstrate our approach on the \ac{METRIXS} beamline at \ac{BESSY II}~\cite{Pietzsch2018} and use the simulation tool RAY-UI \cite{Baumgaertel2019} as a digital twin. As a proof of concept we test our approach not on the real-world beamline but on a beamline simulated with RAY-UI using imitated random offsets. This approach enables rapid testing with known offsets and prevents technical issues like unknown incorrect motor directions leading to wrong signed mirror alignments.

\section{Related Work}
\label{sec:related_work}
Since beamline optimization is a frequent task in all synchrotron radiation sources, numerous attempts are made to optimize automatically the beamline parameters. One approach is based on genetic algorithms~\cite{Xi2017}. This approach is significantly restricted in the range of beamline parameter intervals and in the maximum amount of 15 optimized beamline parameters. Furthermore, only the parameters of two mirrors and one monochromator are optimized.
Karaca et al.~compare several optimization methods: genetic algorithms, non-dominated sorting genetic algorithm II, particle swarm optimization, and artificial bee colony~\cite{Karaca2024}. While particle swarm optimization achieves the best results in this study, its significance is limited, as only the focal distances of two mirrors are optimized.
Another method that is applied is Bayesian optimization~\cite{Morris2024}. 
These experiments are successfully applied for up to eight beamline parameters. Although the problem space is constrained, scaling Bayesian optimization to higher dimensions remains challenging due to its exponentially increasing computational complexity.

In summary, existing beamline optimization approaches are constrained by the number of optimized beamline parameters or are restricted to narrow parameter intervals. These restrictions are required to limit the dimensionality and optimization problem space. All described approaches require parameter evaluations at the actual beamline, as no digital twin exists or offsets compensation is not applied to this digital twin. Determining fixed parameters like mirror radii or slope errors is impossible without matching a virtual twin to the real-world beamline. Furthermore, the possible evaluation number is limited, as real-world beamline evaluations take several seconds, rendering the process too slow for efficient optimization.

We train a surrogate model similar to Nash et al.~\cite{Nash2022} to avoid this bottleneck. Because our surrogate model is a neural network, it is differentiable and, thus, allows the use of gradient-based optimizers. Neural networks also enable batch-wise evaluations on \acp{GPU}, allowing rapid processing of large numbers of evaluations. Due to these properties our offset finder can find global minima by evaluating millions of beamline parameters per second, using an optimizer similar to a random walker approach. In this study, we do not optimize the beamline parameters but focus on identifying the discrepancies between reality and simulation. Optimizing the beamline parameters based on these discrepancies is then straightforward, as it involves defining the desired target footprints and using the same optimizer as in the offset-finding process after adjusting the found offsets of the fixed parameters in the surrogate model. However, this aspect extends beyond the scope of the current work.

\section{Method}
\subsection{Surrogate Model}
\label{ssec:surrogate}
We use a neural network for the surrogate model that replaces the computationally intense raytracing simulation tool, taking the beamline configurations as inputs and generating only the $x$ and $y$ histograms of the footprints, unlike the replaced simulation tool. To calculate the $x$ and $y$ histograms, we divide the intervals of possible values of $x$ and $y$ into $k$ equally sized bins, counting how many rays are within the specified bin, as shown in \cref{fig:params_to_hist}. We use $k=50$ bins for the $x$ histograms over the interval $[-10, 10]$~mm, and $k=50$ bins for the $y$ histograms over the interval $[-3, 3]$~mm. These intervals cover the majority of rays while maintaining high positional accuracy. Unlike the simulation tool, the surrogate model is specific to the beamline and limited to the parameter ranges as defined in \cref{tab:parameters_metrixs}.

We create a dataset with uniformly random sampled beamline parameters from \cref{tab:parameters_metrixs} to train the neural network. These parameters are simulated using RAY-UI with 100\,000 input rays, and their $x$ and $y$ histograms are stored in this dataset. We refer to each specific input constellation, as described in \cref{tab:parameters_metrixs}, combined with the RAY-UI simulation result, as a sample. We create a large dataset of 20 million samples, as approximately 95\% of the beamline parameter combinations within our selected intervals result in a nearly empty footprint. We aim to ensure the dataset includes a sufficient number of samples---around 1 million---with footprints that contain enough rays for an effective training. The number of required samples could potentially be lower, so further investigation is needed to determine the optimal dataset size. From 80\% of the dataset, we select samples containing at least one ray. From the remaining 20\%, we select all samples including empty footprints. We apply this approach for faster dataloading and to speed up training because the dataset is very imbalanced, with samples containing rays being significantly rarer than those without rays. To further address data imbalance, we divide the data into “good” and “bad” groups of samples, where the former is comprised of almost 1 million samples containing at least ten rays, and the latter is comprised of those containing less. Samples are shown to the network with a ratio of 3:1 good to bad samples. A threshold below ten rays for this classification increases the \ac{MSE} for non-empty histograms, while higher thresholds increase the overall \ac{MSE}. The resulting dataset is split into 80\% training, 10\% validation, and 10\% test data. We min-max normalize all inputs and outputs of the neural network and use an architecture of seven linear layers. We choose a batch size of 32, the exponential learning rate scheduler with a starting learning rate of $10^{-4}$, and a decay rate of $0.999$. All layers, except the last layer, use the Mish activation function. As an optimizer, we use ADAM. These hyperparameters are selected through systematic trials.

\subsection{Simulating the Real-World Beamline}
\label{ssec:simulating}
Since testing on a real-world beamline is time-intensive, we test instead on another digital twin. By adding fixed but random offsets to all beamline parameters, this new digital twin imitates the fixed deviations of the real-world beamline from the digital twin. The offset finder’s task is to find offsets so that the footprints of the “real” beamline match those of the digital twin with the same configurations. A \textit{configuration} refers to a specific set of all beamline parameter positions. Various configurations must be scanned to find the offsets as accurately as possible. Since the real-world beamline is fully motorized, these configurations can be scanned automatically and quickly in the final real-world application.

As previously mentioned, configurations with high ray counts are manually identified on the real-world beamline for meaningful results. To imitate this procedure, we use the surrogate model to search for random beamline configurations and add random offsets. We repeat this process until offsets are found for which the sum of the histogram values exceeds the value $0.5$ for $n > 15$ beamline configurations, ensuring sufficient ray counts similar to those from the real-world beamline. As configuration candidates, 1000 real-valued vectors consisting of all beamline parameters are drawn uniformly distributed from the intervals as in \cref{tab:parameters_metrixs}. In the real-world beamline fixed but unknown parameters are chosen randomly but with the same values for all configurations. Furthermore, 100 real-valued offsets are drawn uniformly distributed with a maximum size of $\pm 20\%$---the maximum expected offsets derived from the experience of previously manually identified offsets---of the beamline parameter intervals. Both variables are chosen uniformly distributed to select them as unbiased as possible. The offsets are added to the configurations, evaluated with the surrogate model, and returned if their ray counts are sufficient.

\begin{table}
    \centering
    \caption{Parameters and ranges for the RAY-UI simulations. Values are chosen randomly from a uniform distribution within the specified intervals. Parameters in italics are fixed but unknown.}
    \label{tab:parameters_metrixs}
\footnotesize
\begin{tblr}
{width=\textwidth, rowhead = 1, colspec={lX}}
\hline
\textbf{Component} & \textbf{Parameter [Range]} \\
\hline
Undulator U41\_318eV & translation~X,~Y: $[-0.25, 0.25]$~mm, rotation~X,~Y: $[-0.05, 0.05]$~mrad \\
Slit ASBL & totalWidth: $[1.9, 2.1]$~mm, totalHeight: $[0.9, 1.1]$~mm, translation~X,~Y: $[-0.2, 0.2]$~mm \\
Mirror M1\_Cylinder & \textit{radius:} $[174.06, 174.36]$~mm, rotation~X: $[-0.25, 0.25]$~mrad, Y,~Z: $[-1, 1]$~mrad, translation~X,~Y: $[-1, 1]$~mm \\
Spherical Grating & \textit{radius:} $[109741, 109841]$~mm, rotation Y: $[-1, 1]$~mrad, Z: $[-2.5, 2.5]$~mrad \\
Exit Slit & totalHeight: $[0.009, 0.011]$~mm, translation Z: $[-150, 150]$~mm, rotation Z: $[-0.3, 0.3]$~mrad \\
Mirror E1 & \textit{longHalfAxisA:} $[20600, 20900]$~mm, \textit{shortHalfAxisB:} $[300.72, 304.72]$~mm, rotation~X: $[-1.5, 1.5]$~mrad, Y: $[-7.5, 7.5]$~mrad, Z: $[7, 14]$~mrad, translation Y,~Z: $[-1, 1]$~mm \\
Mirror E2 & \textit{longHalfAxisA:} $[4325, 4425]$~mm, \textit{shortHalfAxisB:} $[96.16, 98.16]$~mm, rotation~X: $[-0.5, 0.5]$~mrad, Y: $[-7.5, 7.5]$~mrad, Z: $[22, 32]$~mrad, translation Y,~Z: $[-1, 1]$~mm \\
\hline
\end{tblr}
\end{table}

\begin{figure}[H]
 \begin{minipage}{\linewidth}
      \centering
      \begin{minipage}{0.49\linewidth}
          \begin{figure}[H]
              \includegraphics[width=\linewidth]{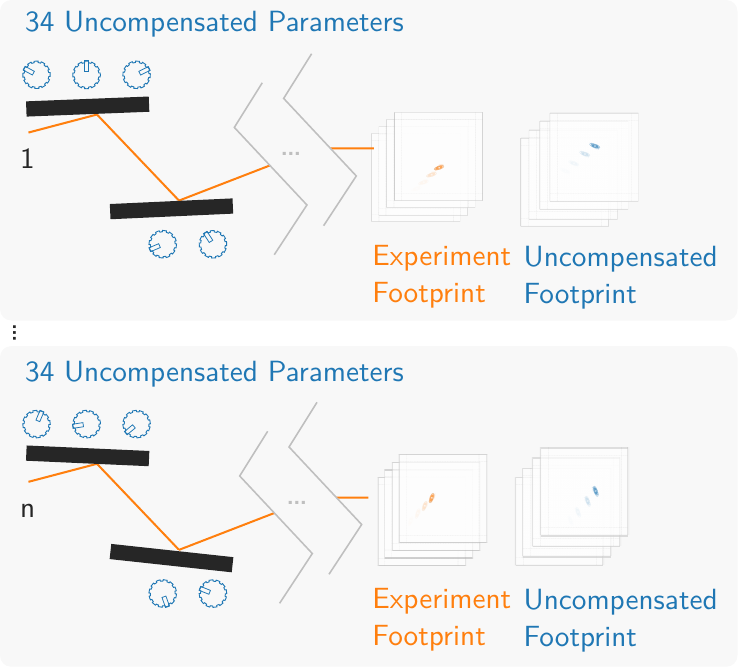}
              \caption{Schematic view of $n$ configurations of the beamline components and its footprints.}
              \label{fig:beamline_description}
          \end{figure}
      \end{minipage}
      \hspace{0.001\linewidth}
      \begin{minipage}{0.49\linewidth}
          \begin{figure}[H]
              \includegraphics[width=0.94\linewidth]{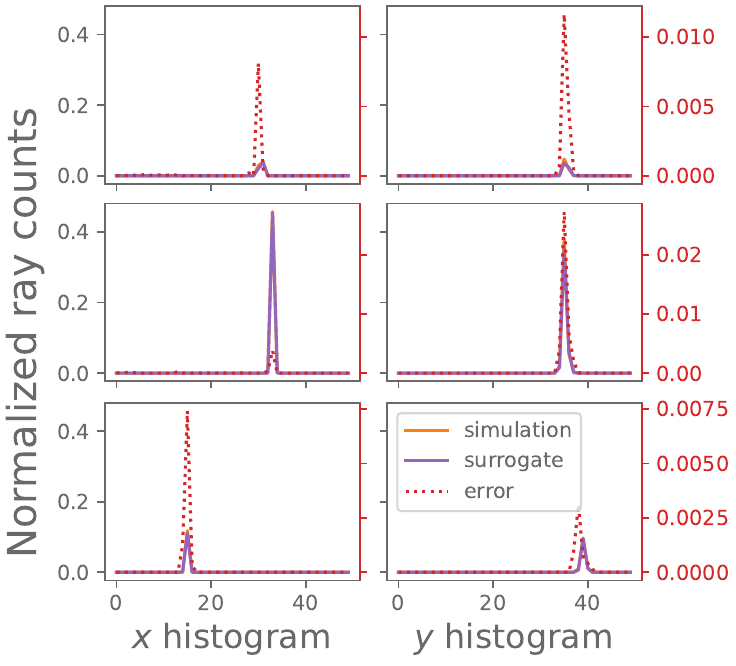}
              \caption{Three histograms of RAY-UI simulation (orange), surrogate model outputs (purple), and absolute error (red).}
              \label{fig:histograms_surrogate_simulation}
          \end{figure}
      \end{minipage}
  \end{minipage}
\end{figure}

\subsection{Offset Finder}
In \cref{fig:beamline_description}, we illustrate the set of $n$ beamline configurations with high ray counts from the simulated real-world beamline. The \textit{experiment footprints} in orange represent the observed rays at the simulated real-world beamline. It is set up with \textit{uncompensated parameters}, represented by the positions of the blue knobs on the left. Simulating the uncompensated parameters in RAY-UI yields the \textit{uncompensated footprints}, depicted in blue on the right. The experiment and uncompensated footprints differ since the experiment footprints are generated with the uncompensated parameters deviated by the generated offsets. Parameters adjusted by the optimizer are called \textit{compensated parameters}, and their simulation in RAY-UI produces the corresponding \textit{compensated footprints}. The target is to find offsets that shift the uncompensated parameters so that the compensated footprints match the experiment footprints precisely. To speed up optimization, the offset finder repeatedly uses the surrogate model instead of the digital twin to evaluate proposed offsets and compare the resulting histograms with those of the experiment footprints.

The offset-finding algorithm, which we will refer to as \textit{smart walker}, works as follows:
\begin{enumerate}
    \item Initialize a fixed amount of $m=1\,000\,000$ offsets with uniformly distributed random values, all below the maximum offsets of 20\% as chosen in the simulated real-world beamline. This choice of $m$ maximizes the number of evaluated solutions while fitting within video memory. 
\item Evaluate these offsets with all $n$ beamline configurations and take only the best solution. For evaluation, we add the offsets to the $n$ beamline configurations and evaluate these offset-adjusted configurations with the surrogate motel, choosing the best solution with the lowest \ac{MSE} between the returned histogram by the surrogate model and the histograms of the observed experiment footprint. We calculate the \ac{MSE} by averaging over the squared errors of all entries of the histograms. 
\item Shift the best solution by an $m$-dimensional, normally distributed vector with standard deviation $\textrm{step\_width}=0.002$, which yields best results in our experiments. The offsets are clamped within their defined ranges. The maximum offsets are set to 20\%, following our choice for the simulated real-world beamline. The shifted solution vector is the next vector of offsets to be evaluated.
\item Continue with step 2 until the maximum number of iterations is reached, which is chosen large enough, ensuring the error between the beamline and surrogate model output is sufficiently small.
\end{enumerate}
We compare our method with a Monte Carlo approach that tries random, uniformly distributed offsets within a maximum range of 20\%, following our choice of the maximum simulated offset. We use the same number of attempted offsets per iteration for comparability as in the smart walker approach.

\section{Results \& Discussion}
\cref{fig:histograms_surrogate_simulation} compares the surrogate model and simulation outputs using the same parameters, showing minimal differences and thus accurate estimations of the ray distributions. Since RAY-UI contains randomly sampled elements, e.g., the exact positions of the rays at the light source, its outputs vary slightly, even though the simulation is started with equal parameters. Thus, a completely perfect estimation is impossible. To compare the outputs of the surrogate model and the simulation, we calculated the overall \ac{MSE} over all samples in the test set and the \ac{MSE} of samples containing at least one ray. The overall \ac{MSE} is calculated on the balanced dataset as described in \cref{ssec:surrogate}. After 300 epochs, an overall \ac{MSE} of $1.77 \times 10^{-6}$ and for the non-empty histograms an \ac{MSE} of $2.16 \times 10^{-6}$ is reached. This result is expected since the samples with rays contain higher weighted values due to the \ac{MSE}.
\begin{figure}
 \begin{minipage}{\linewidth}
      \centering
      \begin{minipage}{0.46\linewidth}
          \begin{figure}[H]
              \includegraphics[width=\linewidth]{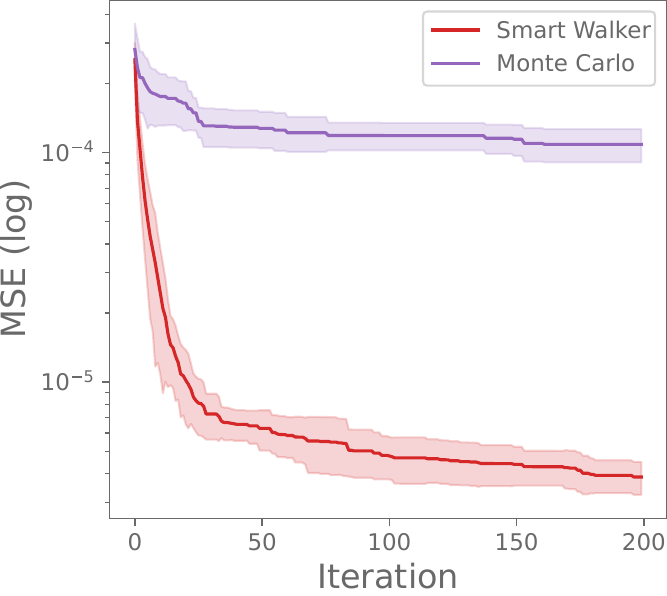}
              \caption{Mean (lines) and standard deviation (shaded) of optimization curves for the smart walker and Monte Carlo methods over ten runs.}
              \label{fig:bl_optimizer_iterations}
          \end{figure}
      \end{minipage}
      \hspace{0.001\linewidth}
      \begin{minipage}{0.49\linewidth}
          \begin{figure}[H]
              \includegraphics[width=\linewidth]{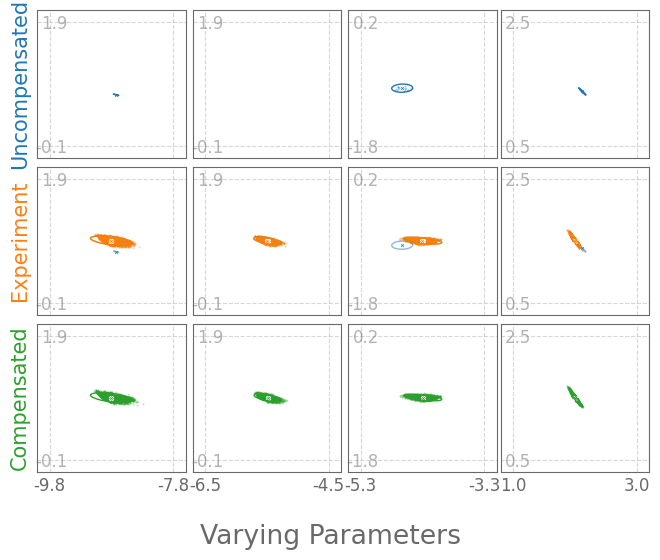}
              \caption{Simulated in Ray-UI: Uncompensated footprints (blue), experiment footprints (orange), compensated footprints (green).}
              \label{fig:fixed_position_plots}
          \end{figure}
      \end{minipage}
  \end{minipage}
\end{figure}

\cref{fig:bl_optimizer_iterations} depicts the optimization curves of the two compared approaches. The smart walker reaches its optimum within 200 epochs and achieves an average \acp{MSE} of $3.87 \times 10^{-6}$ for the histograms of observed and compensated footprints over ten repeated runs, with a standard deviation of $6.34 \times 10^{-7}$. The Monte Carlo approach plateaus early, with an optimum at $1.09 \times 10^{-4}$, and its standard deviation is with $1.79 \times 10^{-5}$ higher and only slightly decreases. The smart walker approach takes, on average, 47.24 seconds on an Nvidia A100 for 200 million evaluations. By that, our approach is more scalable and efficient for beamline optimization, overcoming the dimensional and computational constraints of existing methods. \cref{fig:fixed_position_plots} depicts the smart walker results after 1000~iterations, where the rows show the footprints using uncompensated parameters (top), uncompensated parameters with random offsets (middle), and compensated parameters (bottom). The observed and compensated footprints match almost perfectly, suggesting that the offset finding works, despite being limited to the surrogate model and the histograms of the footprints. The compensated parameters of the smart walker approach achieve an average Sinkhorn distance of all ray coordinates of $3.06 \times 10^{-5}$ compared to the reference with known offsets, while the Monte Carlo approach reaches $7.48 \times 10^{-4}$. For reference, ten repeated runs with the digital twin of the known offsets differ by $2.03 \times 10^{-6}$ from the reference run due to simulation randomness.

The normalized parameters the optimizer finds deviate with a \ac{RMSE} of $0.58$ from the pre-defined, randomly chosen normalized offsets. These high deviations occur because one component’s settings can counterbalance another component’s settings. For example, two mirrors on opposite sides can counterbalance their movements.

\section{Conclusion \& Outlook}
We have trained a highly accurate surrogate model capable of identifying offsets for 34 beamline parameters, allowing us to find offsets up to $\pm$20\% of the beamline parameter intervals. Since only some parameters can be adjusted, we determine and apply the offsets to the digital twin before optimizing beam properties like the smallest possible focus. We must consider $x$, $y$, and $z$ translation errors in the measurement chamber to apply this approach to a real-world beamline. With $z$ translation, we mean a shift in the direction of the beam perpendicular to the $xy$ plane. An extended surrogate model is needed for $z$ translation, while $x$ and $y$ translations can be reconstructed by shifting the histograms. Improving the optimizer through methods like simulated annealing or a weighted solutions population, as in evolutionary algorithms, is necessary to handle the increased problem dimensionality. We can also use gradient-based techniques like stochastic gradient descent for local search because the surrogate model is derivable.

\section{Acknowledgments}
Support by the JointLab AIM-ED between Helmholtz-Zentrum f\"ur Materialien und Energie, Berlin, and the University of Kassel is gratefully acknowledged. We also acknowledge the valuable feedback from the co-authors, Kristina Dingel and Franz Götz-Hahn.

\bibliographystyle{unsrt}  
\bibliography{3_optimization}
\end{document}